\documentclass{aa}

\usepackage{graphicx}
\DeclareGraphicsExtensions{.eps,.ps}
\usepackage{txfonts}
\usepackage{natbib}
\usepackage{amssymb}
\bibpunct{(}{)}{;}{a}{}{,} 

\usepackage{ifpdf}

\ifpdf
\usepackage{epstopdf}
\usepackage[colorlinks=true,linkcolor=magenta,citecolor=dgreen,breaklinks=true]{hyperref}

\fi

\def\target{UGC\,11763}
\def\arcsecpoint{$''\!.$}
\def\arcminpoint{$'\!.$}

\def\pn{EPIC-pn} 
\def\mos{EPIC-MOS}
\def\rgs{RGS}

\newcommand{\kms}{\mbox{km\,s$^{-1}$}}
\newcommand{\Msol}{\textrm{M}\ensuremath{_\odot}}

\newcommand{\fluxunits}{\mbox{erg\,s$^{-1}$\,cm$^{-2}$}}
\newcommand{\funitsa}{\mbox{erg\,s$^{-1}$\,cm$^{-2}$\,\AA$^{-1}$}}
\newcommand{\funitsk}{\mbox{erg\,s$^{-1}$\,cm$^{-2}$\,keV$^{-1}$}}

\newcommand{\plusmin}[2]{\mbox{$^{+#1}_{-#2}$}}

\usepackage{color}

\definecolor{dgreen}{rgb}{0,.5,.1} 
\definecolor{pink}{rgb}{.9,.2,.5}  
\definecolor{orange}{rgb}{.9,.4,0} 
\definecolor{darkred}{rgb}{.545,0.0,.0}
\begin{document}

\titlerunning{UGC\,11763}
\authorrunning{M.V. Cardaci et al.}
\title{Characterization of the emitting and absorbing media around the nucleus
  of the active galaxy UGC\,11763 using XMM-Newton data}

\author{M.V. Cardaci \inst{1,2},  M. Santos-Lle\'o \inst{2}, 
        Y. Krongold \inst{3}, G.F. H\"agele \inst{1,4},  
        A.I. D\'iaz \inst{1}, P. Rodr\'iguez-Pascual \inst{2}} 

\offprints{M.V. Cardaci}

\institute{1 Universidad Aut\'onoma de Madrid, Ctra.\ de Colmenar Km.15,
              Cantoblanco, 28049 Madrid, Spain \\ 
           2 XMM-Newton Science Operations Center, ESAC, ESA, POB 78, E-28691
              Villanueva de la Ca\~nada, Madrid, Spain \\
           3 Instituto de Astronom\'ia, Universidad Nacional Aut\'onoma de
              M\'exico, Apartado Postal 70-264, 04510 M\'exico DF, M\'exico \\
	   4 Facultad de Cs.\ Astron\'omicas y Geof\'isicas, Universidad
              Nacional de La Plata, Paseo del Bosque s/n, 1900 La Plata, 
              Argentina \\
           \email{Monica.Cardaci@sciops.esa.int}
          }

\date{ }

\abstract
{}
{The detailed analysis of all data taken by the {{\em XMM-Newton}}
  satellite of \target\ to characterize the
  different components that are emitting and absorbing radiation in the
  vicinity of the
  active nucleus.}
{The continuum emission was studied through the EPIC spectra taking profit of
  the spectral range of these cameras. The high resolution RGS spectra were
  analyzed in order to characterize the absorbing features and
  the emission line features that arise in the spectra of this source.}
{A power law with a photon index $\Gamma = 1.72^{+0.03}_{-0.01}$ 
  accounts for the continuum emission 
  of this source in the hard X-rays from 10 down to 1\,keV. At lower energies,
  a black 
  body model with $kT= 0.100\pm 0.003$\,keV provides a good description
  of the observed soft 
  excess. The absorption signatures in the spectra of \target\ are consistent
  with the presence of a two phase ionized material 
  ($\log U=1.65^{+0.07}_{-0.08}; 2.6\pm 0.1$ and 
  $\log N_{\mathrm H} = 21.2\pm 0.2$; $21.51\pm 0.01$\,cm$^{-2}$,
  respectively) in the line of  
  sight. The physical conditions found are consistent with the two
  phases being in pressure equilibrium. The low ionization component is
  more ionized than typically found for warm absorbers in other
  Seyfert\,1 galaxies. There are also signatures of some emission lines: 
  O{\sc vii}
  He$\alpha$(r), O{\sc vii} He$\alpha$(f), a blend of the Ne{\sc ix} He$\alpha$
  triplet and  Fe{\sc xviii} at $\lambda\,17.5$\,\AA.} 
{}


\maketitle

\section{Introduction \label{sec:intro}}

\medskip

Narrow Line Seyfert\,1 (NLS1) galaxies are active galactic nuclei that share
many properties with 
Seyfert 1 galaxies, such as a strong continuum and strong 
Fe{\sc ii} emission lines, but whose line widths are similar to those of
Seyfert 2 galaxies \citep{1985ApJ...297..166Osterbrock}.
The more striking characteristics of these objects concern their X-ray
spectra, that show strong soft excess emission 
\citep{1996A&A...305...53Boller}, a rapid and 
large-amplitude variability \citep{2000NewAR..44..387Boller}, and generally
a steeper hard X-ray continua than `normal' Seyfert 1 objects
\citep{1997MNRAS.285L..25Brandt}. 
Different models have been suggested to describe the nature of these objects
\citep{2004ApJ...607L.111Ghosh} and the most accepted paradigm is that NLS1s
possess low mass black holes (about $10^7$\Msol) that are accreting
material close to the Eddington rate 
\citep{2002ApJ...565...78Boroson,2004ApJ...606L..41Grupe}. 

Photoionized X-ray absorbing gas is observed associated with a good number of
NLS1 galaxies  
\citep[see e.g.\ ][ and references therein]{2000NewAR..44..483Komossa}. 
This highly ionized material,
called ``warm absorber", has also been detected in about 50\% of type 1
Active Galactic Nuclei (AGNs), both Seyfert 1s
\citep{1982PhDT.........4Halpern,1997MNRAS.286..513Reynolds,1998ApJS..114...73George}
and quasars \citep{2005A&A...432...15Piconcelli}.  
\cite{1999ApJ...516..750Crenshaw} and \cite{2002ASPC..255...69Kriss} found
that all the objects in their sample 
that exhibit signatures of X-ray warm absorbers also show intrinsic
ultra-violet (UV) absorption. In some cases there is evidence suggesting that
the same medium is responsible for the absorptions in both X-rays and UV 
\citep[e.g.\ ][]{1994ApJ...434..493Mathur,1995ApJ...452..230Mathur,2000ApJ...538L..17Kriss}.
In others, however, this correspondence is not so clear. In NGC\,7469 
\citep{2003A&A...403..473Kriss} and NGC\,3783 \citep{2003ApJ...583..178Gabel}
it was found that some of the UV absorbing components
could be related with the X-ray absorbing ones. On the other hand, in
NGC\,4051, e.g., the physical conditions found for both absorbing components
were different \citep{2009A&A...496..107Steenbrugge}. 
Perhaps due to the complexity of the physics of the absorbing media the exact
relation between the X-ray and UV absorbing systems is still unclear.

\target, also known as II\,Zw\,136, Mrk\,1513 or PG\,2130+099, among other
names, has been classified as a NLS1 galaxy by 
\cite{2003PASP..115..592Constantin}. The width of the H$\beta$ line in this
object is between 2250-2800\,\kms
\citep{1992ApJS...80..109Boroson,2004AJ....127..156Grupe,2008MNRAS.385...53Mullaney,2008ApJ...688..837Grier},
only slightly above the upper limit for the NLS1 classification.
%
%
It was included in the NLS1 sample of \cite{1992ApJS...80..109Boroson} and
\cite{2006MNRAS.368..479Gallo} although \cite{1996A&A...305...53Boller}
excluded it from their list. 
On the other hand, \cite{2006A&A...455..773Veron} have classified it 
as an intermediate Seyfert (Seyfert\,1.5).
\cite{1999ApJS..121..287Huchra} estimated an optical redshift
z\,=\,0.062977 for this object whose position coordinates are 
$\alpha_{\rm 2000}$=21$^h$\,32$^m$\,27\fs81, 
$\delta_{\rm 2000}$=+10$^{\circ}$\,08\arcmin\,19\arcsecpoint46
\citep{1981MNRAS.197..829Clements}.

\target\ has been studied over a wide range of wavelengths.
Using optical spectra 
\cite{2008MNRAS.385...53Mullaney} fitted the two strongest Balmer Hydrogen
lines with three emission line components, the widest of them (FWHM of about
4500 and 5100\,km\,s$^{-1}$ for H$\alpha$ and H$\beta$, respectively) being
typical of the broad lines found in `normal' Seyfert\,1 galaxies, i.e.\
$\sim$\,3000\,km\,s$^{-1}$. Besides, they also estimated the widths of
a broad component in the [Fe\,{\sc vii}]\,$\lambda$\,6087\,\AA\ and 
[Fe\,{\sc x}]\,$\lambda$\,6374\,\AA\ lines ($\sim$\,2000 and
$\sim$\,2700\,km\,s$^{-1}$, respectively) which turned out to be very similar
to the intermediate width component of the Balmer lines ($\sim$\,2100 
and 2350\,km\,s$^{-1}$ for H$\alpha$ and H$\beta$, respectively).
For the central black hole mass, \cite{2004ApJ...613..682Peterson} derived a
value of $4.57\pm0.55 \times 10^{8}$\,\Msol. \cite{2008ApJS..177..103Ho}
reduced this value by a factor of 1.8
($2.5 \times 10^{8}$\,\Msol) for consistency with the virial mass zero point
adopted by \cite{2005ApJ...630..122Greene}.
This value is in the upper limit of the NLS1 black hole mass distribution
found by \cite{2004ApJ...606L..41Grupe}. Nevertheless, it has been further
reduced to $3.8\pm 1.5 \times 10^7 \Msol$ by \cite{2008ApJ...688..837Grier}
more in the range of what is observed in this kind of objects. 

In X-rays, it was observed by {\em EXOSAT}, {\em Einstein}, 
{\em Ginga} and {\em ROSAT} satellites.
\cite{1991ApJ...372...49Singh} analyzed {\em EXOSAT} data to investigate
the variability and the soft excess in the X-ray spectrum. They found that in
one year the
X-ray flux varies by $\sim$\,35\,\% in the 1.5-6\,keV band and by a factor of
$\sim$\,2 in the low energy range (0.1-2\,keV).
They also reported that the low-energy component is very steep
($\Gamma$\,$\approx$\,6) and dominates the spectrum below 0.5\,keV,
irrespective of whether the object is in a high or low state. A soft excess
was also found by \cite{1992A&A...253...35Masnou} using 
{\em Einstein Imaging Proportional Counter}, covering the energy range 
$\sim$\,0.15-3.5\,keV.
  
\cite{1992ApJ...389..157Williams} reported spectra in the 2-20\,keV  range from
{\em Ginga}, and \cite{1997MNRAS.288..920L}, using the same observations,
derived the parameters of the Fe K$\alpha$ emission line at
6.4\,keV. Recently, \cite{2007ApJ...662..860Inoue} studied the \pn\ data from
{\em XMM-Newton} satellite as part of a wider Fe K$\alpha$ line studio. These
authors found that this line is narrow
($\sigma$\,=\,0.02$^{+0.22}_{-0.02}$\,keV) and has an equivalent width (EW) of
139$^{+495}_{-139}$\,eV. 

Regarding the AGN surrounding medium, extreme UV observations with FUSE have
shown absorption features 
due to H (Ly$\alpha$ to Ly$\zeta$), C{\sc iii}\,$\lambda$\,977\,\AA\ and O{\sc
vi}\,$\lambda \lambda$\,1032,1038\,\AA\, originated in an associated system
\citep{2003ApJS..146....1Wakker} whose relative velocity with respect to the
AGN (--1600\,km\,s$^{-1}$) does not allow to rule out that this
absorption system be associated with \target, rather than an intergalactic
cloud.
\cite{2007AJ....134.1061Dunn}, using the same and newer observations from
FUSE, also found this blue shifted component, with a velocity of
--1500\,km\,s$^{-1}$, together with a second one of 20\,km\,s$^{-1}$, both
relative to the AGN. According to these authors, the former component is
clearly visible and virtually 
free of ISM interference, while the latter one is only appreciable in
the O{\sc vi} red member, the blue member of this line being contaminated
with an Fe{\sc ii} and two H$_2$ close lines. These two absorption
components were also found by \cite{1999ApJ...516..750Crenshaw} using 
{\em GHRS-HST} (Goddard High-Resolution Spectrometer-Hubble Space Telescope)
UV spectra.

In this paper we present a detailed analysis of all the available data
obtained with the {\em XMM-Newton} satellite in order to characterize the
circumnuclear environment of this NLS1 galaxy.
In Section \ref{sec:obs} we describe the observations and data
reduction. We present the optical-UV results in Section \ref{sec:opt-uv} and
the X-ray spectral analysis in Section \ref{sec:x-rays}. In Section
\ref{sec:variability} we analyze the source variability during the 
{\em XMM-Newton} observation and compare our observed fluxes with those from
the literature.
Our results are discussed in Section \ref{sec:discussion}. Finally, the summary
and conclusions of this work are given in Section \ref{conclusions}. 


\section{Observations \label{sec:obs}}

\target\ was observed with {\em XMM-Newton} \citep{2001A&A...365L...1Jansen}
on May 16th, 2003. The complete observation, with Id. number 0150470701,
lasted for 39\,ksec. In order to avoid potential pile-up problems, 
EPIC-pn 
and MOS1 
cameras \citep{2001A&A...365L..27Turner} were used in their small window mode
with the 
thin and thick filters, respectively. 
EPIC-MOS2 was operated in full frame mode (thin filter)
in order to allow investigation of other serendipitous X-ray sources in the
field. 
The Reflection Grating Spectrometers
\citep[RGSs,][]{2001A&A...365L...7DenHerder} were run in the default Spectroscopy
mode. 
For the Optical Monitor (OM) \citep{2001A&A...365L..36Mason} we decided to
combine broad-band imaging filters to investigate the circumnuclear structure
in the UV domain with a series of UV-Grism exposures to obtain UV spectral and
variability information of the active nucleus. 
The OM was always operated with windows defined by us: 
we used the largest imaging un-binned window, 5\arcminpoint 1$\times$
5\arcminpoint 0, centered on the target for the broad-band filters and the
default grism window for the UV exposures. Table\,\ref{tabobs} lists
instrument, mode, filter and scheduled exposure time for each instrument.

The data have been processed with the 
7.0.0 version of the Science Analysis Subsystem (SAS) software package
\citep{2001A&A...365L...1Jansen}, 
using the calibration files available on March, 2007. 
All the standard procedures and screening criteria have been followed
in the extraction of the scientific products. 

During most of the observing time the background count rate in the EPIC
cameras was well below 10\% of the source count rate. 
High background time intervals have been excluded
using the method that maximizes the signal to noise in the spectrum as
described in \citet{2004MNRAS.351..161Piconcelli}. 
Hence, the maximum count rate allowed for the background in
`good' periods was 0.5\,c\,s$^{-1}$, 0.5\,c\,s$^{-1}$ and 0.7\,c\,s$^{-1}$ for
\pn, \mos1 and \mos2, respectively. 
For \rgs1 and \rgs2, periods with count rates higher 
than 0.2\,c\,s$^{-1}$ have been excluded. 
The last column of Table\,\ref{tabobs} lists the final exposure time  
after taking into account live time\footnote{The live time is 
the ratio between the time interval during which the CCD is collecting X-ray
events (integration time, including any time needed to shift events towards the
readout) and the frame time (which in addition includes time needed for the
readout of the events). XMM-Newton Users Handbook 
({\tt http://xmm.esac.esa.int/external/xmm\_user\_support/
documentation/uhb/node28.html}).} and high
background screening, per exposure.

\begin{table}[htbp]
\begin{minipage}[t]{\columnwidth}
\caption{Details of {\em XMM-Newton} instrument exposures.}
\label{tabobs}
\centering
{\tiny
\begin{tabular}{lllcc}
\hline\hline
\noalign{\smallskip}
Instrument & Mode & Filter & Time (s) & Time (s) \\
           &      &        & Scheduled & Effective \\
\hline 
\noalign{\smallskip}
EPIC-pn   & Small Window & Thin         & 37511 & 24609$^{\mathrm{a}}$ \\
EPIC-MOS1 & Small Window & Thick        & 37660 & 36476 \\
EPIC-MOS2 & Full Frame   & Thin         & 37672 & 34114 \\
RGS1      & Spectroscopy & --           & 37917 & 36141 \\
RGS2      & Spectroscopy & --           & 37913 & 36145 \\
OM        & Image        & U            &  1000 & \\
OM        &              & UVW1         &2$\times$1000 & \\
OM        &              & UVM2         &2$\times$1000 & \\ 
OM        &              & UVW2         &2$\times$1000 & \\
OM        &              & UV Grism     &19$\times$1000 & \\
\noalign{\smallskip}
\hline 
\noalign{\smallskip}
\end{tabular}}
\end{minipage}
$^{\mathrm{a}}$The live time of the EPIC-pn small window mode is 0.71.
\end{table}


\section{Optical-UV Analysis \label{sec:opt-uv}}

OM data have been processed with SAS task {\sc omichain} and default
parameters. In the broad-band OM images there is no clear evidence of extended
emission. Unfortunately, the spectrum recorded in each individual UV-grism
exposure is very weak; in addition, a number of
zero-order images near the spectrum location and along the dispersion
further complicate the extraction of the source spectrum. As a result,
the total signal-to-noise of the extracted spectra was not as high as
expected. 

The broad-band OM images provide flux measures in the U, UVW1, UVM2 and UVW2
filters. 
For the last three bands, two consecutive exposures are available; the
flux differences in these three pairs of consecutive exposures are all
compatible with no variability within the measurement errors ($<$3\%). 
Table\,\ref{om-fluxes} shows the effective wavelengths of these filters
together with measured fluxes.

\begin{table}
\begin{minipage}[t]{\columnwidth}
\caption{Fluxes in the OM filters obtained using aperture photometry.}
 \label{om-fluxes}
 \centering
 \begin{tabular}{lccc}
  \hline\hline
  \noalign{\smallskip}
  Filter & Effective  wavelength  & Flux \\
         & (\AA)                  & ($10^{-14}$\funitsa) \\
  \hline 		     
  \noalign{\smallskip}	     
  U      & 3440                   & 1.05$\pm$0.01 \\
  UVW1   & 2910                   & 1.24$\pm$0.01$^{\mathrm{a}}$ \\
  UVM2   & 2310                   & 1.24$\pm$0.02$^{\mathrm{a}}$ \\
  UVW2   & 2120                   & 1.58$\pm$0.05$^{\mathrm{a}}$ \\
  \noalign{\smallskip}
  \hline 
  \noalign{\smallskip}
\end{tabular}
\end{minipage}
$^{\mathrm{a}}$ Fluxes are the mean values of the two
  exposures on each filter.
\end{table}

\begin{figure}
\resizebox{\hsize}{!}{\includegraphics{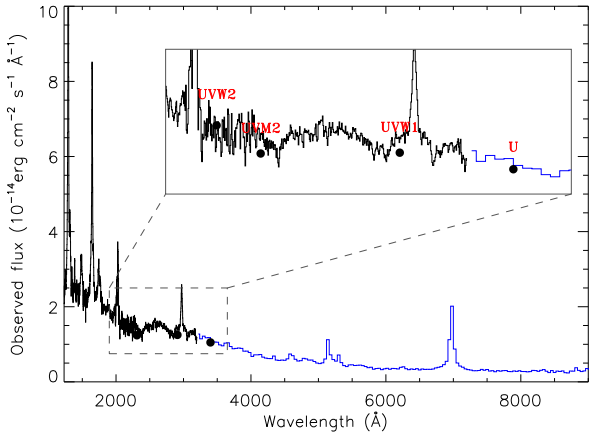}}
 \caption{Average IUE spectrum of \target\ (1200-3200\,\AA, in black) merged
 with the optical spectrum (3200-9000\,\AA, in blue) from
 \citet{1978AJ.....83.1257DeBruyn}. 
 Solid red circles show the OM measures for the different filters.} 
 \label{figuvspec}
\end{figure}

\target\ was observed by the {\em International Ultraviolet Explorer} (IUE)
in six different epochs between 1978 and 1985, with its Short Wavelength
(1150--1950\AA) and Long Wavelength (1950--3200\AA)
spectrographs. Variations of $\sim$25\%\ around the mean are found
along the whole IUE range, with the ratio of maximum to minimum UV flux being
close to 2. OM flux measurements are 
overplotted in Fig.~\ref{figuvspec} on a merged spectrum constructed by
combining the average IUE 
spectrum and the optical one presented by
\cite{1978AJ.....83.1257DeBruyn}.

The difference between the OM measures and the average IUE spectrum at the
effective wavelengths of the OM filters is consistent with the combined errors
of the IUE average spectrum and the OM filter sensibility. 
Therefore, we
conclude that the average UV and the optical spectra are indeed an
acceptable representation of the UV-optical spectral energy distribution (SED)
of \target\ at the time of the {\em XMM-Newton} observation.
From the IUE average spectrum we take the UV flux at 2500\,\AA,
$F(2500\AA)=1.53\times 10^{-14}$\,\funitsa, that we have used to compute the
optical/X-ray spectral index $\alpha_{ox}$
\citep{1979ApJ...234L...9Tananbaum}.
 This value has neither been corrected for the Balmer
continuum nor the Fe{\sc ii} contributions.


\section{Variability \label{sec:variability}}

\begin{figure}
\centering
\includegraphics[angle=0,width=0.44\textwidth]{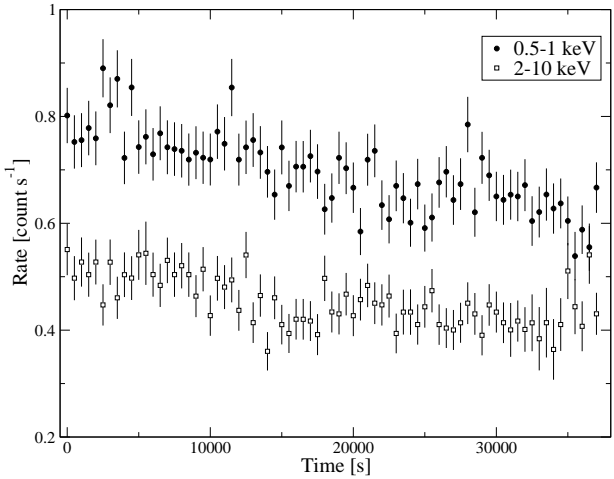} 
\caption{Soft (filled circles) and hard (open squares) \pn\ light
    curves binned by 500\,s.}
 \label{lc}
\end{figure}

We have analyzed the \pn\ soft (0.5-1\,keV) and hard (2-10\,keV) X-ray
background subtracted light curves (Fig.\,\ref{lc}) to investigate the
variability of this source during the observation by computing the statistical
validity of constant flux assumption. Values of $\chi^{2}_{\nu}$ of 2.2 and 1.3 are found for the soft and hard energy bands respectively. Taking into account that a
$\chi^{2}_{\nu}$$>$1.2 corresponds to a probability of less than 10\%
and $\chi^{2}_{\nu}$$>$1.4 to a probability of less than 1\% for the
data to be well represented by a constant flux value, this implies that the 
flux varied during the observation. 

The overall behavior of the light curves seems to show a decrease
of the flux from the beginning to the end of the observation. 
Applying a linear fit, we find a rate of change in the count rates of about 
$-5.1(\pm 0.5)\times 10^{-6}$ counts\,s$^{-2}$ ($\chi^{2}_{\nu}=0.9$) 
and $-2.7(\pm 0.5)\times 10^{-6}$ counts\,s$^{-2}$ ($\chi^{2}_{\nu}=0.8$)
for the soft and hard bands, respectively. 
The statistic of the linear fit, with probabilities of 73 and 86\% for the
data to be well represented by the linear model, implies that this simple model
only reproduces the general trend of the flux variation. 
The maximum flux decrease during the observation can be quantified by
computing the ratio between the maximum and the minimum rate which is found to
be $1.7^{+0.3}_{-0.2}$ and $1.5^{+0.4}_{-0.3}$ for the soft and hard X-ray
bands, respectively.

We have compiled several values
of X-ray fluxes published for \target\ along the years (Table\,\ref{x-flux}). 
This object was also serendipitously observed by {\em XMM-Newton} (\pn)
during a slew on the 14th of May 2006. We have also included in
Table\,\ref{x-flux} the soft and hard X-ray fluxes as provided in the
XMM-Newton slew survey Source Catalogue
\citep[XMMSL1.2,][]{2008A&A...480..611Saxton}. 
\target\ shows a
large amplitude variation in its soft X-ray flux. If we compare the flux in
the 0.1-2.4\,keV range obtained from the {\em ROSAT} observation with our
measurement we find about an order of magnitude variation between these two
epochs. 
The 2006 data of the slew observation show that \target\ had returned to
flux values similar to those of 1990.
In the 2-10\,keV range the amplitude of the flux variation is higher
than a factor of 2.
Unfortunately, there is no data in the hard spectral
range corresponding to the observed highest soft X-ray flux. 
Taking into account the fluxes in Table\,\ref{x-flux}
for this object, 
it becomes clear that, in 2003, {\em XMM-Newton} observed it in the lowest
activity 
state so far reported.

\begin{table}[htbp]
\begin{minipage}[t]{\columnwidth}
\begin{center}
\caption{Comparison between X-ray fluxes from the literature and from
  this work.}
\label{x-flux}
\begin{tabular}{@{}lccccc@{}}
\hline\hline
\noalign{\smallskip}
Data & Obs.\ date & \multicolumn{3}{@{}c@{}}{Flux}  & Ref. \\
\cline{3-5}
 & & \multicolumn{2}{@{}c@{}}{Soft} & \multicolumn{1}{@{}c@{}}{Hard} & \\
 & & 0.1-2\,keV & 0.1-2.4\,keV & 2-10\,keV & \\
\noalign{\smallskip}
\hline	
\noalign{\smallskip}			  

EXOSAT & 1984 Nov.\ & 4.2 &  --  & 5.6  & 1 \\
EXOSAT & 1985 Nov.\ & 9.5 &  --  & 6.3  & 1 \\
GINGA  & 1989 Nov.\ & --  &  --  & 4.7  & 2 \\
ROSAT  & 1990 Nov.\ & --  & 33.1 & --   & 3 \\ 
XMM    & 2003 May   & --    &  --  & 3.63 & 4 \\[1pt]
XMM    & 2003 May   & ${2.70_{-0.08}^{+0.19}}$ & $3.00\pm 0.08$ & ${3.65_{-0.08}^{+0.07}}$ & 5 \\[2pt]
XMM    & 2006 May   & ${16\pm 2} ^{a}$ & -- & ${10\pm3} ^{b}$ & 6 \\

\noalign{\smallskip}			  
\hline 
\noalign{\smallskip}			  
\end{tabular}
\end{center}
\end{minipage}
Fluxes are given in $10^{-12}$\,\fluxunits. 
$^{a}$Flux between 0.2-2\,keV. $^{b}$Flux between 2-12\,keV. 
References: (1) \cite{1991ApJ...372...49Singh}; 
(2) \cite{1992ApJ...389..157Williams}; (3) \cite{1999A&A...349..389Voges}; 
(4) \cite{2007ApJ...662..860Inoue}; (5) This work; 
(6) XMM-Newton slew survey Source Catalogue \citep{2008A&A...480..611Saxton}. 
\end{table}

\section{X-ray spectral analysis \label{sec:x-rays}}

All the EPIC data were checked for no pile-up using the SAS task 
{\sc epatplot}. 
The coordinates of the center of the EPIC-pn X-ray source are 
within 1$''$ of the optical coordinates of \target\ quoted in 
\S\ref{sec:intro}.
The EPIC spectra were extracted using the standard parameters.
For the EPIC-pn spectrum the source extraction area was a circular region of
32\arcsecpoint5 in radius. 
The background extraction area was a circular
region at 2\arcminpoint0 
North-NW of the target and of 32\arcsecpoint5 in radius.
The \mos1\ source spectrum was extracted using a circular region of 40$''$
radius. The background region was an annulus centered in the source with
8$'$ and 9\arcminpoint2 inner and outer radius, respectively. 
The \mos2\ source spectrum was extracted  using a circular region with 1$'$
radius and the background was extracted from an annular region also centered
in the target position and 2\arcminpoint5 and 3\arcminpoint6 inner and outer
radius, respectively. 
The three EPIC spectra have been rebinned to have at least 30
counts per bin and a 
maximum of three channels per resolution (FWHM) element of the detector.

The RGS data has been processed with the SAS task {\sc rgsproc} and default
parameters, except that we asked for the background subtracted spectra. 
In spite of \target\ being a bright object when observed by {\em ROSAT} 
\citep{1999A&A...349..389Voges}, 
it seems to be in a lower state when observed by {\em XMM-Newton}
satellite (see discussion in \S\ref{sec:variability}). As a result,
the scheduled time for the {\em XMM-Newton} observation does not allow to
achieve as good signal-to-noise in the RGS spectra as expected. We therefore
had to bin the spectra, loosing resolution, but increasing the
signal-to-noise ratio. 
The RGS spectra were geometrically binned to avoid any smoothing
of the absorption 
and emission features. We chose a 15 channels binning as a
compromise between loosing resolution and having the necessary spectral
signal-to-noise for a statistically significant fit. 
Hence, as the default spectral bin size is 10\,m\AA, at 15\,\AA, our final
spectra have bins of about 150\,m\AA\ at that wavelength. 

All spectra have been fitted using {\em Sherpa} package of {\em CIAO 3.3}
\citep{2001SPIE.4477...76Freeman}. We use the $\chi^2$ statistics with the
Gehrels variance function \citep{1986ApJ...303..336Gehrels} and the Powell
optimization method. 
The first because it is based on Poisson
statistics for small number of counts in a bin and on Binomial statistics
otherwise, and the second because is a
robust direction-set method for finding the nearby fit-statistic minimum.

\subsection{Low resolution spectra \label{lr-spectra}}
\begin{figure}
\resizebox{\hsize}{!}{\includegraphics[angle=-90]{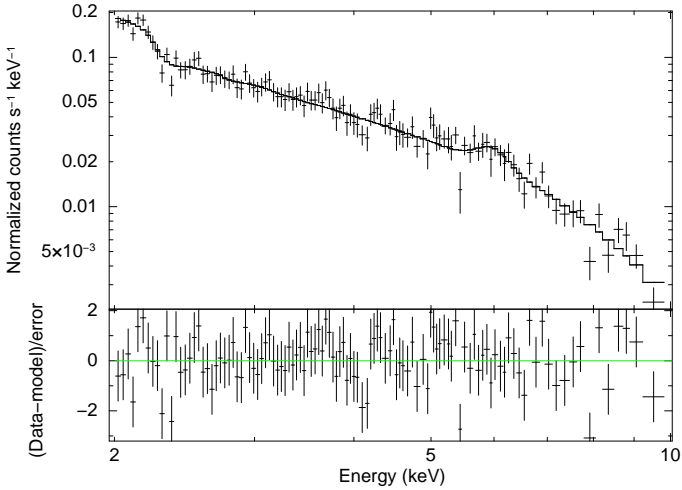}}
 \caption{EPIC-pn spectrum of the UGC\,11763 nucleus, showing the power law 
 fit to the 2-10~keV range. Galactic absorption and a neutral Fe-K$\alpha$
 fluorescence line are also included in 
 the fit. (See Table\,\ref{tabpn} for the best fitting parameters.)}
 \label{pn-2-10}
\end{figure}

An absorbed power law has been fitted to the \pn\ spectrum in the
2.0-10.0~keV energy range. The H column density has been fixed to the
Galactic value of  $N_\mathrm{H}=4.67\times10^{20}\,{\mathrm{cm}}^{-2}$ 
\citep{1990ARA&A..28..215Dickey}. 
The result (see Table\,\ref{tabpn}) has a reduced $\chi^2$ of 
$\chi^{2}_{\nu}$=1.07 for 112 degrees of freedom ({\it dof}). 
Adding a redshifted Gaussian emission line
at the energy of the neutral Fe-K$\alpha$ fluorescence line does
improve the fit (Table\,\ref{tabpn}), with a new value of
$\chi^{2}_{\nu}=0.99$ for 109 {\it dof} (F=4.1, probability
99.19\%). 
The line is weak with an equivalent width of
0.23$^{+0.15}_{-0.11}$\,keV. 
The model parameter values are
listed in Table\,\ref{tabpn} and the fit is shown in Fig. \ref{pn-2-10}.
Leaving the absorbing column density as a free parameter 
the neutral H column density takes a value lower than the Galactic one,
indicating that a neutral absorber, either local or at the \target\ redshift,
does not improve the fit. 
Hence, unless otherwise specified, all the models mentioned from here on
include neutral absorption by the Galaxy with the column density fixed to
above value. 

\begin{table}[htbp]
\begin{minipage}[t]{\columnwidth}
\begin{center}
\caption{EPIC-pn fit model parameters for the 2-10~keV energy
  range.} 
\label{tabpn}
\begin{tabular}{lll}
\hline\hline
\noalign{\smallskip}				  
\parbox[c]{0.6in}{Model\\component}	& Parameter               & Value 	    \\
\noalign{\smallskip}
\hline	
\noalign{\smallskip}			  
powerlaw	& $\Gamma$         & $1.65_{-0.05}^{+0.06}$  \\[3pt]
 		& $K_{pwlw}$       & $8.3_{-0.6}^{+0.6}$	    \\[3pt]
statistic       & $\chi^{2}_{\nu}$ &  1.07                   \\
                & $\mathit{dof}$   &  112                    \\
\noalign{\smallskip}
\hline				  
\noalign{\smallskip}
powerlaw	& $\Gamma$         & $1.69_{-0.06}^{+0.06}$  \\[3pt]
		& $K_{pwlw}$       & $8.6_{-0.3}^{+0.4}$	    \\[3pt]
Gaussian	& $E_{rest}$       & $6.35_{-0.34}^{+0.16}$  \\[3pt]
		& $\sigma$         & $0.26_{-0.11}^{+0.52}$  \\[3pt]
		& $K_{line}$       & $0.9_{-0.5}^{+1.1}$     \\[3pt]
statistic       & $\chi^{2}_{\nu}$ &  0.99                   \\
                & $\mathit{dof}$   &  109                    \\
\noalign{\smallskip}
\hline
\noalign{\smallskip}
\end{tabular}
\end{center}
\end{minipage}
The 
  galactic $N_\mathrm{H}$ is fixed to $4.67 \times 10^{20}$ 
  cm$^{-2}$. The line energy in the rest frame of the source ($E_{rest}$) and
  $\sigma$ of the Fe-K$\alpha$ emission line are given in keV; power-law
  normalizations ($K_{pwlw}$) in units of $10^{-4}$
  ph\,\,keV$^{-1}$\,cm$^{-2}$\,s$^{-1}$ at 1 keV; and line
  normalization $K_{line}$ in units of $10^{-5}$ ph\,\,cm$^{-2}$\,s$^{-1}$.
  Errors quoted are at 90\% confidence level.
\end{table}
%
%
When the 1.0-2.0\,keV range is included in the fit (i.e. a 1.0-10.0\,keV \pn\
fit is performed), the results do not change significantly 
indicating little or no effects of any other component in this energy range.

\begin{figure}
\resizebox{\hsize}{!}{\includegraphics[angle=-90]{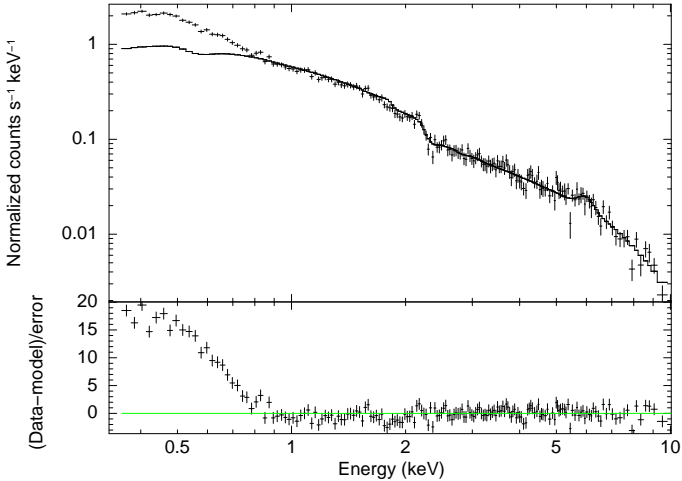}}
 \caption{EPIC-pn spectrum of the UGC\,11763 nucleus, showing the whole,
 0.35-10~keV energy range and the fit to the 2-10~keV range. The fit has
 two components, a power law and a Gaussian emission line at 6.36\,keV, 
 absorbed by the Galactic column density. 
 A huge soft excess below $\sim 0.9$~keV is clearly
 shown in the residuals panel}
 \label{pn-soft-excess}
\end{figure}

\begin{figure}
 \includegraphics[angle=0,width=0.48\textwidth]{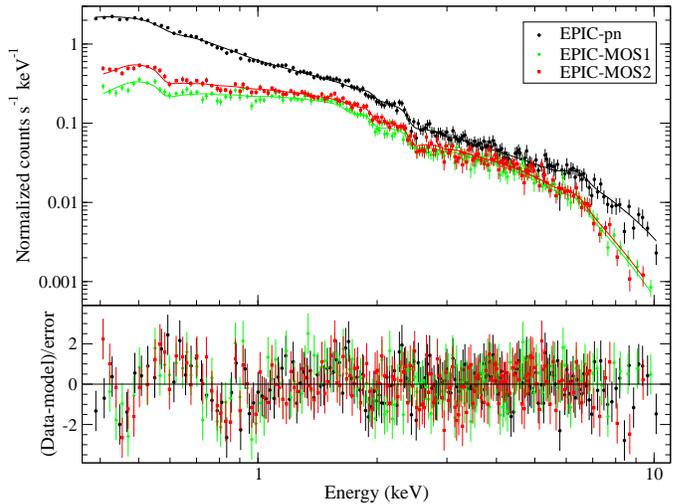}
 \caption{\pn, \mos1 and \mos2 spectra -in the rest frame- of the
   \target\ nucleus, showing the best fit model $\mathcal{A}$
   (Table\,\ref{tablepic}). The model includes Galactic 
   absorption plus a power law, a black body and the Fe-K$\alpha$ line. 
   Residuals between 0.7 and 0.8 keV are clearly seen in the lower panel
   consistent with the presence 
   of absorption by ionized material. See text for more details.}
\label{pn-mos-cont}
\end{figure}

%
%
Plotting the 2-10~keV energy range model overlapped to the whole 0.35-10~keV
energy range (Fig. \ref{pn-soft-excess}) it becomes clear that the flux below
1\,keV exceeds the extrapolation of the harder power law flux. This is a
frequent feature of AGN X-ray spectra and usually referred to as the soft X-ray
excess. 
Typical models used to fit the soft excess are one or several black bodies,
bremsstrahlung radiation or a higher photon index power law
\citep[e.g.][]{2005A&A...432...15Piconcelli}. 
Using these models to describe the three \target\ EPIC spectra in the
0.35 to 10\,keV range, we find that a model with a black body component
(hereafter model $\mathcal{A}$) 
with $kT = 0.09$\,keV provides the best description of the soft excess
observed in this object (Table\,\ref{tablepic}). 
Nevertheless, although the overall fit is good ($\chi^{2}_{\nu}=0.97$ for 509
$\mathit{dof}$), there are large residuals at low energies  
(see Fig.~\ref{pn-mos-cont}).
They are largest around 0.7-0.8\,keV or 16-17\AA, where a blend of
several iron inner M-shell absorption lines, known as the Fe `Unresolved
Transition Array' (UTA), has been detected in other Seyfert\,1 galaxies and
interpreted as the signature of ionized gas in the line of sight
to the X-ray source \citep[e.g.][]{2001A&A...365L.168Sako}. 
There are also large residuals around 0.5-0.6\,keV where some oxygen emission
lines could be present (O{\sc vii}\,$\lambda\lambda$\,21.6, 21.8, 22.1\,\AA).

\begin{table}[htbp]
\begin{minipage}[t]{\columnwidth}
\begin{center}
\caption{Model $\mathcal{A}$ parameters for the simultaneous \pn, \mos1 and
  \mos2 spectrum fit in the 0.35-10~keV energy range. } 
\label{tablepic}
\begin{tabular}{@{}lll@{}}
\hline\hline
\noalign{\smallskip}				  
\parbox[c]{0.6in}{Model\\component}	& Parameter  &  Value             \\
\noalign{\smallskip}
\hline	
\noalign{\smallskip}			  
powerlaw	& $\Gamma$   & $1.63 \pm $\tiny{0.02} 	\\
                & $K_{pwlw}$ & $7.9 \pm $\tiny{0.2}     \\[3pt]
Gaussian	& $E_{rest}$ & $6.36_{-0.27}^{+0.15}$   \\[3pt]
		& $\sigma$   & $0.34_{-0.15}^{+0.33}$ 	\\[3pt]
		& $K_{line}$ & $1.0_{-0.4}^{+0.6}$       \\[3pt]
black body	& $kT$       & 0.090$ \pm $\tiny{0.002} \\
                & $K_{bb}$   & $3.2 \pm $\tiny{0.2}     \\
statistic & $\chi^{2}_{\nu}$ &  0.97                    \\
          & $\mathit{dof}$   &  509                     \\
\noalign{\smallskip}
\hline				  
\noalign{\smallskip}
\end{tabular}
\end{center}
\end{minipage}
The galactic
  $N_\mathrm{H}$ is fixed to $4.67 \times 10^{20}$ cm$^{-2}$. The line energy
  is given in the rest frame of the source. The normalizations
  $K_{pwlw}$ and $K_{bb}$ correspond to the \pn\ spectrum. 
  $E_{rest}$, $\sigma$ and $kT$ in keV,
  $K_{pwlw}$ in $10^{-4}$ ph\,\,keV$^{-1}$\,cm$^{-2}$\,s$^{-1}$ at 1\,keV,
  $K_{line}$ in $10^{-5}$ ph\,cm$^{-2}$\,s$^{-1}$, and $K_{bb}$ in units of
  $10^{-5}$ $L_{39}/D_{10}^{2}$ where $L_{39}$ is the source luminosity in
  units of $10^{39}$ erg\,s$^{-1}$ and $D_{10}$ is the distance to the source
  in units of 10\,kpc. Errors quoted are at 90\% confidence level.
\end{table}

\subsection{High resolution X-ray spectra \label{hr-spectra}}

The better resolution of the \rgs\ spectra as compared to the EPIC spectra
favors 
the identification and fitting of the absorption and narrow emission
features. 
Hence, in order to take advantage of the whole information acquired by
{\em XMM-Newton} we
use simultaneously the \pn, the two \mos\ and the two high resolution \rgs\
spectra.
\pn\ and \mos\ data are restricted from 0.35 to 10\,keV, and
\rgs\ data are taken between 0.41 and 1.8\,keV. 

The residuals of the EPIC fit from model $\mathcal{A}$ seem to
disclose the presence of partially ionized absorbing material in the line of
sight to the source. The signature of this warm material, a characteristic Fe
unresolved transition array (UTA), is clearly seen on the high resolution RGS
spectra.

To get a good identification of the soft X-ray spectral features and
determine the physical properties of the absorbing gas, we have made use of
the PHASE (PHotoionised Absorption 
Spectral Engine) photoionization code \citep{2003ApJ...597..832Krongold}. 
The parameters that are usually let free are the ionization parameter of the
gas, the equivalent hydrogen column density and the outflow
velocity. Another  
parameter is the internal micro-turbulent velocity of the gas, which is
very difficult to constrain because single absorption lines in the spectrum
are unresolved by the actual available instruments and most of the
observed features are blends of several transitions. Therefore 
we have fixed this velocity to 300\,\kms, the same value used by
\cite{2003ApJ...597..832Krongold} and \cite{2001ApJ...554..216Kaspi} to fit
the spectra of NGC\,3783. 
We note, however, that the exact value of this parameter has little effect on
the results \citep[see][]{2009ApJ...690..773Krongold}.
The ionization parameter is defined as the ratio between the
density of ionizing photons and the density of hydrogen particles,  
$U=Q(H)/4\pi c r^2 n_{\mathrm H}$, where $r$ is the distance to the source,
$n_{\mathrm H}$ is the hydrogen number density, $c$ the speed of light and
$Q(H)$ is the rate of hydrogen ionizing photons (or the integral over all
hydrogen-ionizing photons of the ratio between the luminosity $L_\nu$ and the
energy in the same frequency, 
$Q(H)=\int_{\nu=13.6eV}^{\infty}{L_\nu/h\nu\, d\nu}\big)$. 

\begin{figure}
 \centering
 \includegraphics[angle=0,width=0.4\textwidth]{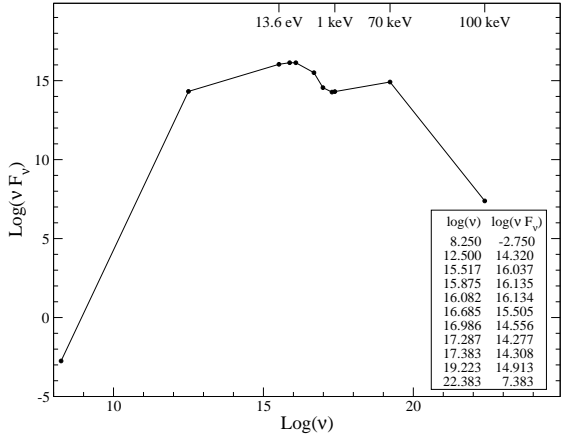}
 \caption{Spectral energy distribution adopted for \target. Values in the
 optical-UV are estimated using the OM data and for energies higher than
 $\sim 0.1$\,keV we have used the \pn\ continuum model.}
\label{sed-ugc}
\end{figure}

The intrinsic SED of the source (Fig.\,\ref{sed-ugc}) is used
to calculate, with CLOUDY 
\cite[version 08,][]{1998PASP..110..761Ferland},
a grid of photoionization models to build the input table for
PHASE. 
The optical-UV part of the SED is obtained from the OM data and for energies
higher than $\sim 0.1$\,keV we have used the \pn\ continuum model.
In this way, the adopted SED represents the emission of \target\ at the
observation time.  
Only the photons with energies higher than $\sim 0.1$\,keV will affect
the absorption in the X-ray spectral range 
\citep{2001ApJ...554..216Kaspi,2003ApJ...597..832Krongold}.  
Nevertheless, the optical-UV part of the SED is important when we calculate
the thermal equilibrium curve discussed in \S\ref{sec:discussion} below.

The fits with PHASE confirm the presence of warm material absorbing the source
X-ray emission. Model $\mathcal{B}$ is the same as $\mathcal{A}$ but including
one PHASE component. 
The result of the fitting using model~$\mathcal{B}$ (Table\,\ref{model.abs})
indicates that the absorbing gas is ionized (T=$1.9\pm 0.3 \times 10^{5}$\,K), 
with an ionization parameter 
log\,U=1.7$_{-0.1}^{+0.4}$ and has a moderate column density
log\,$N_{\mathrm H}$=21.0$\pm 0.2$\,cm$^{-2}$ ($\chi^{2}_{\nu}$= 0.97,
$\mathit{dof}$= 719). The F-test shows that, on the EPIC spectra, the fit with
model $\mathcal{B}$ is better than with model $\mathcal{A}$ 
with a 99.9\% of confidence level.
The estimated value of the redshift of the absorbing material is 
0.062$\pm 0.001$, to be compared to a measured optical redshift of 0.063 for
\target\ (\S\ref{sec:intro}). This, therefore may indicate that the
material is flowing out from the nucleus with a velocity of about
$300\pm300$\,\kms, 
but it is also compatible with material at rest relative to the nucleus.

The \rgs2 residuals (at about 11-12 \AA) suggest the existence of a second
absorber. Therefore we have considered a new model with two absorbers
(Model 
$\mathcal{C}$, Table\,\ref{model.abs}).  
The F-test shows that the fit with two warm absorbers is better than the fit
with only one absorber at 99.9\%  confidence level. 
The second absorption component is compatible with it being highly ionized
gas,  
(log\,U=2.6$\pm0.1$ and T=$1.2\pm0.1 \times 10^{6}$\,K) with a column density 
log\,$N_{\mathrm H}$=21.51$\pm0.01$\,cm$^{-2}$.
The inclusion of this component modified the parameters of the
other model components: $\Gamma, K_{pwlw}$ and kT are slightly higher than
in Model $\mathcal{B}$, and the U and N$_{\mathrm{H}}$ of the first absorption
component are lower and higher respectively (Table\,\ref{model.abs}). 
There is also a change in the velocity of the low ionization 
absorbing component. For the Model $\mathcal{B}$ the velocity is compatible
with material flowing out or being at rest with the nucleus, but for the 
Model $\mathcal{C}$ both absorbing components have
velocities that are compatible with material flowing
into the nucleus.

Besides the absorption features, there are also emission signatures on the
high resolution spectra. Narrow emission lines of O{\sc viii}-Ly$\alpha$,
O{\sc vii}-He$\alpha$ and Ne{\sc ix}-He$\alpha$, among others, are
frequently found in similar objects
\citep[e.g.][]{2000ApJ...535L..17Kaspi,2003ApJ...594..128Turner,2003A&A...403..481Blustin,2004MNRAS.350...10Pounds,2008A&A...484..311Longinotti},
hence, we have included in our model the emission features as Gaussian line
profiles. The procedure that has been followed is described below.

The lines are added on a one-by-one basis to check their statistical
significance, constraining their energies to vary in a small range around
their laboratory energies ($E_{lab}$) and taking into account the
redshift of the source.
Since we cannot use only the F-test as a reliable
criteria to compute the statistical significance of a Gaussian line
\citep{2002ApJ...571..545Protassov}, we use a combined method to decide
whether to 
include or not a line in our final model. Along with the F-test we check
carefully the fit and the residuals, and finally we also take into account the
wavelength positioning of the line.

\begin{figure*}
\centering
\includegraphics[angle=0,width=.7\textwidth]{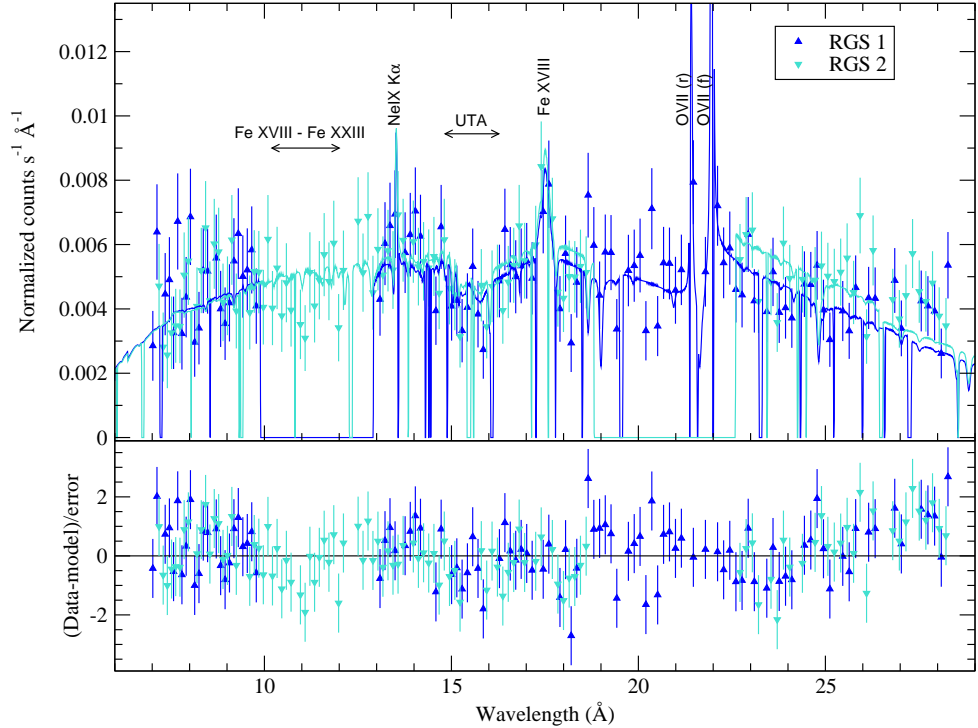}
 \caption{High resolution X-ray spectra -in the rest frame- of \target\ as
 obtained with RGS1, (blue up triangles), and RGS2 (cyan down triangles), and
 binned to 15  channels per bin. Solid lines are the convolution of the best
 fit model with instrument responses. (Model $\mathcal{C}$.) }
 \label{figmodrgs}
\end{figure*}

We first include in the model the O{\sc vii}-He$\alpha$ lines triplet.
The intercombination line is not found in the fitting process
so we take it off and keep only the recombination ($r$) and forbidden ($f$)
lines. The width of these lines takes a very small value (less than
$10^{-5}$\,keV) 
so we have fixed them to zero, i.e.\ to the instrumental resolution. In this
way we find that the $r$ and $f$ lines of the  O{\sc vii}-He$\alpha$ triplet
are statistically significant for the fit (probability of 97\%). These oxygen
lines lie in the range of energies where no RGS2 data are available due to its
non operational CCD, and where there are also bad pixels in the RGS1 spectrum. 
Therefore, their parameters 
are not well constrained (see Fig.\,\ref{figmodrgs} and
Table\,\ref{model.lines}). Nevertheless we consider that they are significant
to the fit. 

The next line added is O{\sc viii}-Ly$\alpha$. This line turns out to be only
marginally significant to the fit (probability of 86\%), and its parameters
are not well constrained either. Moreover, it lies just at the border of the
non-operational CCD of the RGS2 instrument. Therefore we do not keep this line
in our final model. To investigate its possible presence on the \target\
spectra a longer observation is needed.

The emission signature seen at 17-18 \AA\ has been modelled with a Gaussian
line with its central energy varying between 0.62 and 0.73\,keV (17-20\AA). 
The fitting process found that it is broad ($\sigma$ about 1700 \kms) 
and statistically significant (probability of 98\%). 
This line could correspond to Fe{\sc xviii} at $\lambda$\,17.62\AA. 
The Ne{\sc ix}-He$\alpha$ triplet lines lay in the 13-14 \AA\ range.
We have added them to the
model finding that only one line is clearly detected and statistically
significant (probability of 98\%). This line could be a blend of the 
three components. 

\begin{figure*}
\centering
\includegraphics[angle=0,width=.7\textwidth]{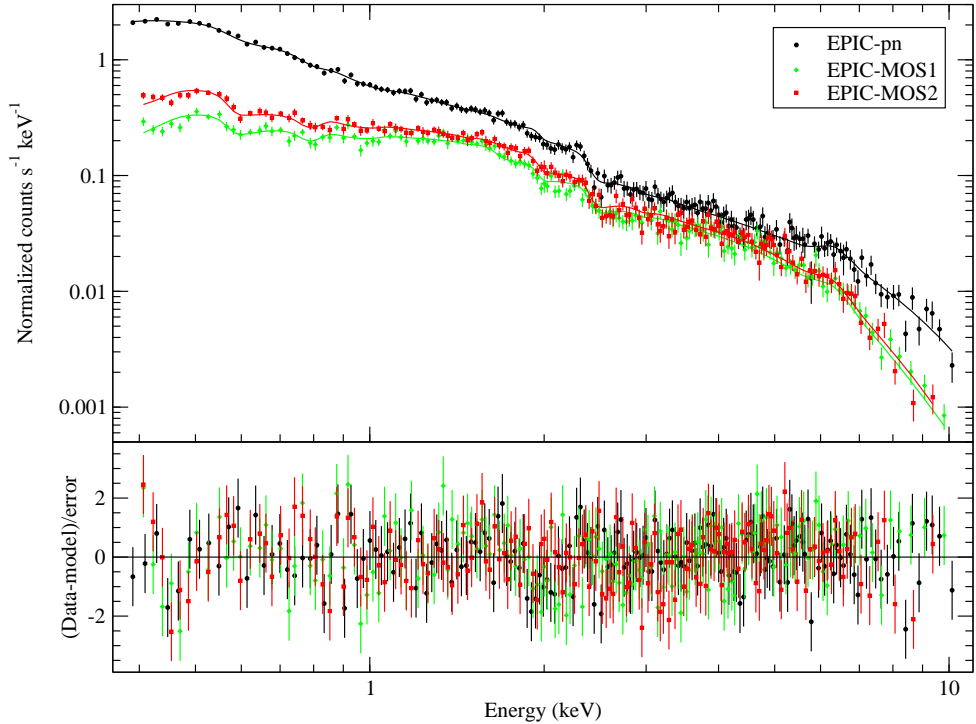}
 \caption{Low resolution X-ray spectra -in the rest frame- of \target\ as
 obtained with the \pn\ (black dots),
 \mos1\ (green diamonds) and \mos2 (red squares) cameras. The best fit model
 (Model $\mathcal{C}$) convolved with the instrument response of each camera
 is shown in solid line.}  
 \label{figmodepic}
\end{figure*}

The parameter values of the best fit model (Model $\mathcal{C}$) are summarized
in Tables\,\ref{model.abs} and \ref{model.lines}. This model and its errors
are plotted in Figs.~\ref{figmodrgs} and \ref{figmodepic}.

\begin{table*}[htbp]
\begin{minipage}[t]{\textwidth}
\begin{center}
\caption{PHASE and continuum parameter values for models $\mathcal{B}$ and
  $\mathcal{C}$ for the simultaneous fits to the EPIC and RGS spectra.}  
\label{model.abs}

\begin{tabular}{@{}l ll @{}l ll @{}l llr @{}l llr l@{}}
\hline\hline
\noalign{\smallskip}
  & \multicolumn{2}{c}{Powerlaw}     
&&  \multicolumn{2}{c}{Black body}             
&& \multicolumn{3}{c}{LIC}         
&& \multicolumn{3}{c}{HIC}   
& \multicolumn{1}{c}{$\chi^{2}_{\nu}$/$\mathit{dof}$}   \\
\cline{2-3} \cline{5-6} \cline{8-10} \cline{12-14}
\noalign{\smallskip}

& \multicolumn{1}{c}{$\Gamma$} & \multicolumn{1}{c}{$K_{pwlw}$} 
&&  \multicolumn{1}{c}{kT} & \multicolumn{1}{c}{$K_{bb}$} 
&& log\,U & log\,$N_{\mathrm H}$ & \multicolumn{1}{c}{vel.}\ 
&& log\,U & log\,$N_{\mathrm H}$ & \multicolumn{1}{c}{vel.} &\\
\noalign{\smallskip}
\hline	
\noalign{\smallskip}			  

model $\mathcal{B}$ & $1.67_{-0.02}^{+0.03}$ & $8.3_{-0.2}^{+0.3}$ 
&& 0.097$_{-0.003}^{+0.003}$     & $3.4_{-0.2}^{+0.2}$ 
&& $1.7_{-0.1}^{+0.4}$ & $21.0_{-0.2}^{+0.2}$    &  $-300_{-300}^{+300}$ 
&&&&
&0.97/719 \\[3pt]
									       
model $\mathcal{C}$ & 1.72$_{-0.01}^{+0.03}$ & $8.9_{-0.3}^{+0.3}$ 
&& $0.100_{-0.003}^{+0.003}$     & $3.4_{-0.3}^{+0.2}$ 
&&  $1.65_{-0.08}^{+0.07}$ & $21.2_{-0.2}^{+0.2}$    & $500_{-300}^{+300}$ 
&& $2.6_{-0.1}^{+0.1}$ & $21.51_{-0.01}^{+0.01}$ & $500_{-300}^{+300}$ 
& 0.90/706 \\

\noalign{\smallskip}
\hline 
\noalign{\smallskip}
\end{tabular}
\end{center}
\end{minipage}
Model $\mathcal{B}$ is similar to model $\mathcal{A}$ with one absorbing
  component and model $\mathcal{C}$ to model $\mathcal{B}$ with another
  absorbing component and a few Gaussian emission lines (see their parameters
  in Table\,\ref{model.lines}). Both $\mathcal{B}$
  and $\mathcal{C}$ models include the Galactic absorption (with fixed 
  $N_{\mathrm H}=4.67\times 10^{20}$\,cm$^{-2}$) and the Fe-K$\alpha$ line
  (with its parameters fixed to the model $\mathcal{A}$ values). 
  Normalizations $K_{pwlw}$ and $K_{bb}$ correspond to the \pn\
  spectrum. 
   $K_{pwlw}$ in units of $10^{-4}$\,ph\,\,keV$^{-1}$\,cm$^{-2}$\,s$^{-1}$; 
   kT in keV; $K_{bb}$ in
   $10^{-5}$\,$L_{39}/D_{10}^{2}$ where $L_{39}$ is the source luminosity in
   units of $10^{39}$ erg\,s$^{-1}$ and $D_{10}$ is the distance to the source
   in units of 10 kpc. $N_{\mathrm H}$ is given in cm$^{-2}$. Velocities, 
   given in \kms, are relative to the systemic velocity of \target.  
  Errors quoted are at 90\% confidence level.
\end{table*}

\begin{table}[htbp]
\begin{minipage}[t]{\columnwidth}
\begin{center}
\caption{Model $\mathcal{C}$ parameters for the narrow emission
  lines. } 
\label{model.lines}
\begin{tabular}{@{}llllll@{}}
\hline\hline
\noalign{\smallskip}
\multicolumn{1}{c}{Line} &\multicolumn{1}{c}{$E_{lab}$} &\multicolumn{1}{c}{$E_{rest}$}&\multicolumn{1}{c}{$\lambda_{rest}$} & \multicolumn{1}{c}{$\sigma$}  & $K_{line}$      \\
\noalign{\smallskip}
\hline	
\noalign{\smallskip}			  
O{\sc vii}           (f) & 0.5610 & $0.564^{(a)}$             & $21.97^{(a)}$           & $0$ (frozen)            & $16_{-14}^{+14}$ \\[3pt]
O{\sc vii}           (r) & 0.5740 & $0.579^{(a)}$             & $21.41^{(a)}$           & $0$ (frozen)            & $5_{-3}^{+3}$  \\[3pt]
Fe{\sc xviii}          & 0.7035 & $0.708_{-0.004}^{+0.005}$ & $17.50_{-0.09}^{+0.11}$ & $0.004_{-(a)}^{+0.004}$ & $3_{-3}^{+1} $ \\[3pt]
Ne{\sc ix}  K$\alpha$    & blend  & $0.917_{-0.015}^{+0.007}$ & $13.52_{-0.22}^{+0.10}$ & 0.0001$^{(a)}$          & $2_{-1}^{+2} $ \\[3pt]
\noalign{\smallskip}
\hline 
\noalign{\smallskip}
\end{tabular}
\end{center}
\end{minipage}
$E_{lab}$ is the laboratory energy of the lines. $E_{rest}$ and 
  $\lambda_{rest}$ are the energy and wavelength of the lines given in the
  rest frame of the object. 
  $E_{lab}$, $E_{rest}$ and $\sigma$ are given in keV, $\lambda$ in
  \AA\ and $K_{line}$ in $10^{-5}$\,ph\,cm$^{-2}$\,s$^{-1}$.   
  $^{(a)}$Unconstrained parameter, see text. Errors quoted are at 90\% confidence level.
\end{table}

\section{Discussion \label{sec:discussion}}

From the OM data of \target\ we have found that the IUE spectra is
representative of its flux 
state during the {\em XMM-Newton} observation.
Using the IUE flux at 2500\,\AA\ and the flux at 2\,keV from
\pn\ data, $F(2\,keV)=7\times 10^{-13}$ \funitsk, we obtain an
optical/X-ray spectral index $\alpha_{ox}=1.6$ for this object. This value
is similar to those given by \cite{1987ApJ...323..243Wilkes} and
\cite{2006MNRAS.368..479Gallo} and is typical of Seyfert\,1 AGNs as shown by
these authors. 

Regarding the shape of the continuum emission from the AGN, a power law
accounts for the hard X-rays. It has a standard spectral index for Seyfert\,1 galaxies or AGNs 
\citep{2005A&A...432...15Piconcelli} and is in
good agreement with the values found by \cite{2007ApJ...662..860Inoue} from
the same {\em XMM-Newton} observation and by
\cite{1991ApJ...372...49Singh} from {\em EXOSAT Observatory} data taken
with the low-energy and medium-energy detectors and the source in a low
luminosity state.
 
The EPIC data show the possible presence of the Fe\,K$\alpha$ line around
6\,keV.
The line parameters are unconstrained
(as seen by the large error bars); in addition, the line is weak (equivalent
width of 0.23$^{+0.15}_{-0.11}$\,keV) and rather broad
($\sigma$=16\,000\plusmin{15\,000}{7\,000} 
\kms). Therefore we consider the significance of this line to be very low.
It is interesting to note that \cite{2007ApJ...662..860Inoue} also included
the Fe\,K$\alpha$ 
line in their final model, at a similar
energy within the errors. Its reported width is smaller than found here, but
their larger $\sigma$ error makes their value and ours compatible. 

The extrapolation of the hard-X-ray model to lower energies, down to
0.35~keV, 
reveals the existence of a large soft-X-ray flux in excess of the hard power
law, from $\sim\,0.9$\,keV and below. 
This excess has also been reported by \cite{1992A&A...253...35Masnou} from 
{\em Einstein Observatory} IPC (Imaging Proportional Counter) data and
\cite{1991ApJ...372...49Singh} from {\em EXOSAT Observatory} data. 
After testing the most typical components used to model the soft excess we
find that it can be accounted for by a black body component with  
$kT\sim 0.1$\,keV (consistent with the value obtained by
\citealt{2007ApJ...662..860Inoue}) as found in many other Seyfert galaxies
\citep{2005A&A...432...15Piconcelli}.

In fact, we find that the continuum spectral characteristics of \target\ are
typical of Seyfert 1 galaxies. 
\target\ shows soft X-ray excess emission as detected in several Seyfert 1 and
NLS1 objects \citep{1996A&A...305...53Boller,2004MNRAS.351..161Piconcelli}.
Regarding variability: in the long term (several years) variations by a factor
of about 10 have been detected; in the short term, variations by about 1.5
have been detected in our 37\,ks observation.
This behavior is typical of `normal' Seyfert 1s 
\citep[e.g.,][]{2005ApJ...635..180Markowitz,2008ApJ...674..686Winter,2008ApJ...689..762Dewangan}.
The rapid and large-amplitude variability as seen in other NLS1s
\citep{2000NewAR..44..387Boller} is not detected during our observation.
For example, for the NLS1 NGC\,4051, \cite{1995MNRAS.273..549McHardy} reported 
a flux variation by a factor of about 10 during a 28\,ks {\em ROSAT}
observation.
However, as can be seen in the middle right panel of Fig. 1 of
\cite{2007MNRAS.375.1479Smith}, NGC\,4051 has also periods during which no
large-amplitude variability is observed. 
The X-ray properties of \target\ do not show the extreme properties of other
NLS1 in X-rays.
Moreover, \cite{2006MNRAS.368..479Gallo} suggests that NLS1s show complex
spectra while at a low flux state. 
We caught \target\ at a flux about 10 times lower than in other epochs. 
Still none of the complex spectral properties attributed by Gallo to NLS1s in
low state have been detected.
We therefore conclude that \target\ is probably a NLS1 at the border between
the NLS1 and 
Seyfert 1 classification. Its optical properties suggest a NLS1 but there is
no conclusive prove of this nature from the present X-ray observations.

Our simultaneous analysis of the EPIC and RGS data reveals the presence of
absorption by ionized gas in the line of sight to \target. 
Two absorbing components are clearly required to fit the data (see
Table\,\ref{model.abs}) as has already been
found for many other objects that show the presence of warm
absorbers (e.g., IRAS 13349+2438, \citealt{2001A&A...365L.168Sako}; 
NGC\,7469, \citealt{2003A&A...403..481Blustin};
NGC\,3783, \citealt{2003ApJ...597..832Krongold}; 
Mrk\,279, \citealt{2007A&A...461..121Costantini};
NGC\,985, \citealt{2009ApJ...690..773Krongold}).
A low ionization component (LIC) -- log\,U=1.65$^{+0.07}_{-0.08}$,
$N_{\mathrm H}$=1.6$\pm0.9\times 10^{21}$\,cm$^{-2}$
(T=$1.8\pm 0.2 \times 10^5$\,K) --, gives rise to absorption by the
Fe M-shell UTA (charge states {\sc xii} to {\sc xvi}).
A high ionization component (HIC) -- log\,U=2.6$\pm0.1$, $N_{\mathrm H}$=3.2$\pm0.1\times 10^{21}$\,cm$^{-2}$, twice higher than for the LIC,
(T=$1.2\pm 0.1 \times 10^6$\,K) -- 
produces absorption by Fe{\sc xx}-Fe{\sc xxv}.
The velocity of both components relative to the systemic
velocity of \target\ is consistent with a single value of
$500\pm 300$\,\kms.

After the inclusion of the second absorber, residuals at 11\,\AA\ on the
RGS2 data (Fig.\,\ref{figmodrgs}) were still found that are consistent with an
absorption 
feature by Fe{\sc xxiii}.
This led us to test the presence of a third absorption component. 
The inclusion of this component in the model, however, does not
improve the fit. 
This result does not rule out the existence of a third absorbing
component in the line of sight to this source, but data with a higher
signal-to-noise ratio are required  in order to establish or discard its
presence.

Different scenarios have been suggested  to model the observed complexity of
the absorbing material in AGNs. 
\cite{2003ApJ...597..832Krongold,2005ApJ...620..165Krongold,2009ApJ...690..773Krongold}
and \cite{2003ApJ...599..933Netzer} modelled the absorption with a discrete
number of absorbing components finding that these components appear to be in
pressure balance with each other.
With a different idea, 
\cite{2003A&A...402..477Steenbrugge,2005A&A...434..569Steenbrugge} and 
\cite{2004ApJ...606..151Ogle} 
suggested that the absorption could be 
produced by a radial distribution of material with a continuous distribution of
temperatures.
However, recent models have shown that models consisting in discrete phases in
pressure balance can also model the ionized absorbers in these objects
(\citealt{2007ApJ...659.1022Krongold}, for NGC\,4051; and
\citealt{2009ApJ...692..375Chelouche},  
for NGC\,5548).
Alternatively, it has also been suggested that, rather than a multi-phase
medium in pressure balance, the absorber could be a single system 
with a constant total (gas plus radiation) pressure with a stratified
distribution in temperature 
\citep{2006A&A...452....1Rozanska,2006A&A...451L..23Goncalves}.

As pointed out by \cite{2003ApJ...597..832Krongold} a wide well
resolved UTA feature sets tight restrictions on the ionization degree of the
absorbing material and different values of the ionization
parameter lead to different shapes and wavelength ranges for the UTA.
Figure\,\ref{figmodrgs} shows a well defined UTA feature in the RSGs spectra of
\target. A medium with a smooth distribution of temperatures is unlikely to
produce this well defined feature 
\cite[for a detailed discussion see][]{2003ApJ...597..832Krongold}. 

\begin{figure}
\centering
\includegraphics[angle=0,width=0.48\textwidth]{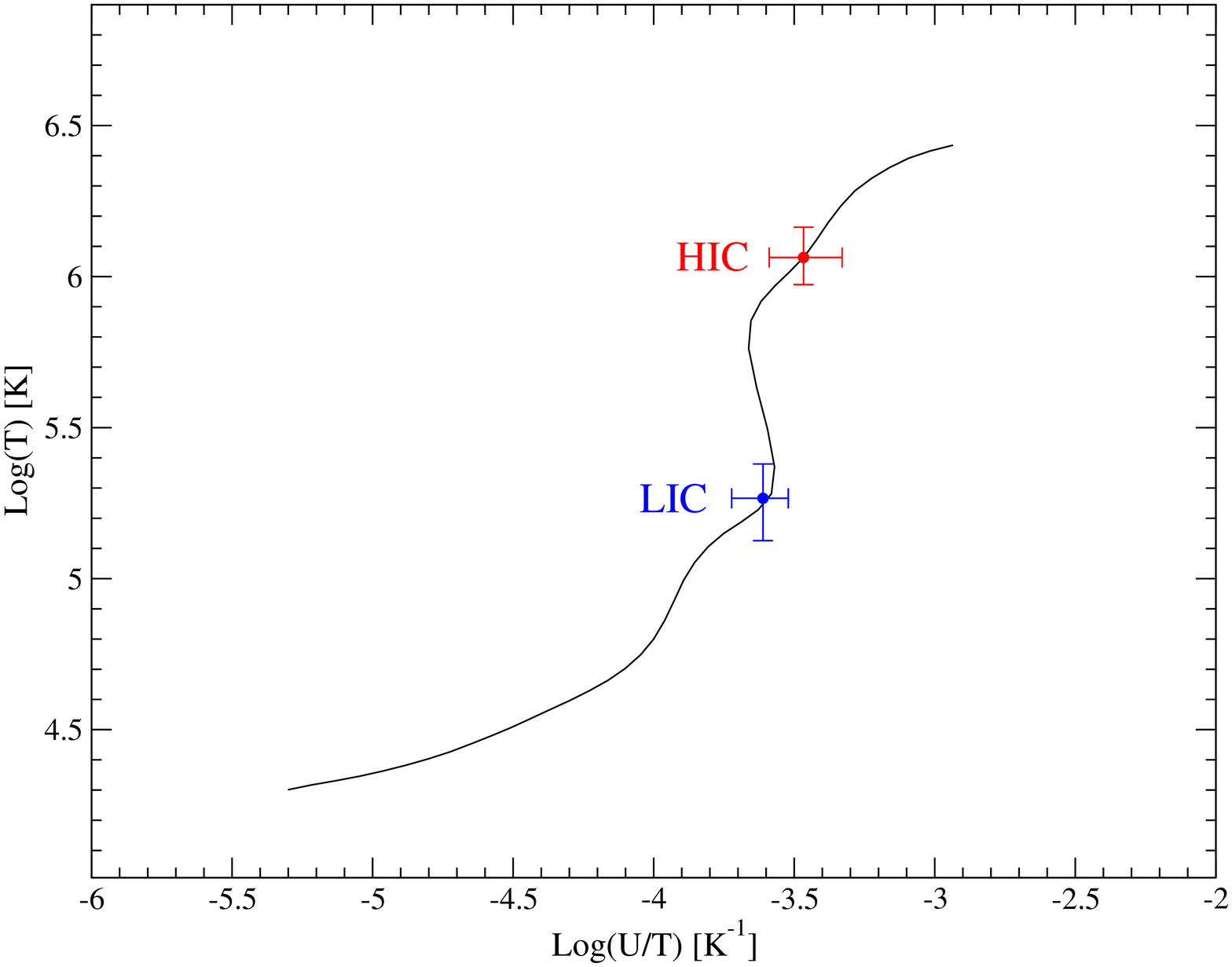}
 \caption{Thermal stability curve of \target\ obtained using the SED based on
 the simultaneous multiwavelength data obtained in this observation and
 assuming a column density of  log\,$N_{\mathrm H}$= 21.5\,cm$^{-1}$. We have 
 indicated the position of the two components of photoionized absorbing
 material from model $\mathcal{C}$. }
 \label{stability}
\end{figure}

Our model presents a simple picture: two kinematically indistinguishable gas
components with rather different photoionization equilibrium temperatures and
two widely separated ionization parameter values. This simple picture suggests
that the absorption observed in the source spectrum may arise from two
phases of the same medium.
To test the possibility of having a multi-phase medium in pressure equilibrium
we have calculated the thermal equilibrium curve 
$\log(T)$ vs.\ $\log(U/T)$ (hereafter S-curve) for \target\
(Fig. \ref{stability}) using the SED described in \S \ref{hr-spectra}. 
The S-curve represents the points where heating and cooling 
processes are in equilibrium. The $\log(U/T)$ value is inversely proportional
to the gas pressure, so that vertical lines in the plot indicate isobaric
conditions. 
More than one phase may exist at pressures where the S-curve is
multivalued  
\citep{2001ApJ...561..684Krolik,2009MNRAS.393...83Chakravorty}. 
Regions of the curve with negative slope are unstable because 
any isobaric perturbation will produce net heating or cooling in the gas,
see \cite{2005ApJ...620..165Krongold} for further details.
For this object we find that the LIC and HIC components lie in stable parts 
of the curve (Fig.\,\ref{stability}).
Both components are consistent with having roughly the same gas pressure. 
This fact and the similar velocity (see Table \ref{model.abs})
we find for both absorbers suggest that the two components in \target\ could
indeed constitute a multi-phase medium. 

In this sense, it is noteworthy that the LIC has a relatively high
temperature ($T=1.8\pm0.2\times10^5$\,K). 
Given the shape 
of the S-curve, only gas at such temperature could coexist in pressure
equilibrium with the high ionization HIC component. 
In fact, the UTA in this object is formed by 
Fe{\sc xii}-Fe{\sc xvi}; 
this is striking since UTAs
found in warm absorbers in other Seyfert 1 galaxies are colder ($T\sim$ few
$\times 10^{4}$\,K) and are produced by Fe{\sc vii}-Fe{\sc xii} 
\citep[see][and references therein]{2009ApJ...690..773Krongold} and in those
objects such charge states are the ones 
required for pressure equilibrium between their absorbing components (for
them, the gas producing Fe{\sc xiii}-Fe{\sc xvi} lies on unstable regions of
the S-curve).

As we have mentioned in \S \ref{sec:intro}, the same AGNs that show
absorption lines in the UV also show signatures of X-ray absorbers
\citep{1995ApJ...452..230Mathur,1999ApJ...516..750Crenshaw,2002ASPC..255...69Kriss}.
In many cases, the absorbers show similar outflow velocities and similar
ionization states in the UV and X-rays, and in this cases, it is
straightforward to 
assume that the same media is producing absorption in both bands 
\citep[e.g.][]{2007ApJ...658..829Arav}. 
Nevertheless, since this is not the case for all the objects studied, the
precise 
relation between the absorbers in both bands is still uncertain.
\cite{1999ApJ...516..750Crenshaw} and
\cite{2007AJ....134.1061Dunn,2008AJ....136.1201Dunn} have studied the UV
spectrum of \target. \citeauthor{1999ApJ...516..750Crenshaw} found signatures
of two 
absorbing components in the lines due to C{\sc iv}, N{\sc v}, and 
O{\sc vi} with radial velocities of $-1568\pm39$ and $+45\pm41$\,\kms. Later,
\citeauthor{2008AJ....136.1201Dunn} also found two components
in the lines due to O{\sc vi}($\lambda\lambda$ 1032, 1038)
with radial velocities similar to those found by 
\citeauthor{1999ApJ...516..750Crenshaw}. 
However, these reported velocities do not agree with the ones we find for the
X-ray absorber.   
Moreover, the
ionization states we find in the LIC and HIC components are too high for 
O{\sc vi} to produce detectable absorption features in the UV \citep[see
for example][]{2007ApJ...658..829Arav}. This implies
that the systems producing the absorption in X-rays are
not the same as those producing absorptions in the UV.

\begin{figure}
\centering
\includegraphics[angle=0,width=0.48\textwidth]{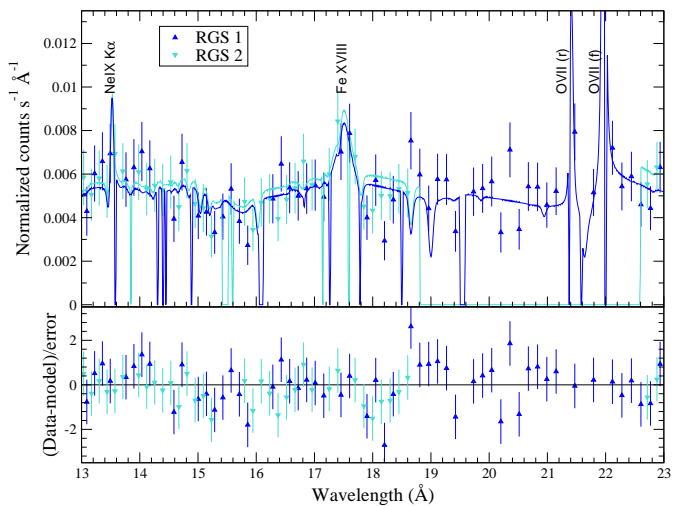}
 \caption{Enlargement of Fig.\,\ref{figmodrgs} showing the fitted lines. }
 \label{lines}
\end{figure}

We find four statistically significant emission lines (see
Fig.\,\ref{lines}). Three of them can be identified as O{\sc vii}
He$\alpha$(r), O{\sc vii} He$\alpha$(f) and a blend of the Ne{\sc ix}
He$\alpha$ triplet. 
The energies of the lines are slightly different from their laboratory
values. Nevertheless, taking into account the errors in the energy
determination 
of the lines during the fitting process, their positions are
consistent with the rest frame of \target.
These lines are often found in 
other Seyfert\,1 and NLS1 AGNs as NGC\,3783 \citep{2000ApJ...535L..17Kaspi}, 
NGC\,3516 \citep{2003ApJ...594..128Turner}, 
NGC\,4051 \citep{2004MNRAS.350...10Pounds} and 
Mrk\,335 \citep{2008A&A...484..311Longinotti} among others.
The fourth line can be identified as the Fe{\sc xviii} line at 17.62\,\AA,
also detected in the spectra of NGC\,7469 by \cite{2007A&A...466..107Blustin}.
None of these lines are resolved in the spectra, which does not allow  the
reliable determination of their widths. Therefore, it is possible that all of
them arise in the same medium. 

As a final remark we would like to point out that both warm absorbing
components appear to be 
redshifted ($z=0.0647\pm 0.0009$) with respect to the emission lines
and the rest frame of the object. This could be indicating the presence of an
inflow. However, 
given the loss in spectral resolution we have to assume in
order to increase the signal-to-noise of our spectra, 
the quality of the data is not sufficient to constrain the 
velocity of these components. 
While the presence of infalling ionized gas could have important potential
effects for our understanding of AGN winds, the present measurements are not
conclusive. A longer observation of this source is needed to further study
this possibility, as well as to constrain the physical parameters of the
emitting media. 


\section{Summary and conclusions  \label{conclusions}}

We have analyzed all the data of \target\ taken by the {\em XMM-Newton}
satellite.  
Joining the optical-UV information with the X-ray information we have
built the SED of this object at the epoch of observation.
We have concentrated our efforts mainly on the analysis of the X-ray
spectra, finding that the continuum emission of the source can be
characterized by a power law and a black body components. The continuum
emission is absorbed by ionized material in the line of sight to the source.
We have found two absorbing components that are consistent with their being in
pressure equilibrium, indicating that they could constitute two
phases of the same medium. This idea is supported by an UTA of higher
ionization than those found in other AGNs.
Furthermore, we have found some emission lines, among them an unusual 
Fe{\sc xviii} emission line. 
The X-ray properties of \target\ described here are compatible with it
being either a normal Seyfert 1 galaxy or a NLS1.  There is
no sign of the extreme X-ray properties, like large amplitude short-term
variability or spectral complexity, detected in other NLS1 at least at
some epochs.
More data are needed in order to get a full characterization of this object.

\medskip

\paragraph{Acknowledgments}
We are grateful to the referee, Elisa Costantini, for a 
careful revision of this manuscript that greatly improved its clarity.
This research is completely based on observations obtained with 
{\em XMM-Newton}, an
ESA science mission with instruments and contributions directly funded by ESA
Member States and NASA. For the spectral fitting use of software
provided by the Chandra X-ray Centre (CXC) in the application package Sherpa
has been made. 

This work has been supported by DGICYT grants AYA-2004-02860-C03 and
AYA2007-67965-C03-03. GH and MC
acknowledge support from the Spanish MEC through FPU grants AP2003-1821 and
AP2004-0977. Furthermore, partial support from the Comunidad
de Madrid under grants S-0505/ESP/000237 (ASTROCAM) and 
S-0505/ESP-0361 (ASTRID) is acknowledged.
MC and GH also thank the hospitality of the UNAM.
YK acknowledges support from the Faculty of the European Space Astronomy
Centre (ESAC), and the hospitality of ESAC.

\bibliographystyle{aa}
\bibliography{ugc11763-cardaci}
\end{document}